\documentclass{webofc}

\usepackage[varg]{txfonts}   
\usepackage{hyperref}
\usepackage{url}
\hypersetup{colorlinks=true,citecolor=blue,urlcolor=blue,linkcolor=blue}
\begin{document}
\title{
Machine learning approach to QCD kinetic theory }
%
%

\author{\firstname{Sergio} \lastname{Barrera Cabodevila}\inst{1}\fnsep\thanks{\email{sergio.barrera.cabodevila@usc.es}} \and
        \firstname{Aleksi} \lastname{Kurkela}\inst{2} \and
        \firstname{Florian} \lastname{Lindenbauer}\inst{3,4}
}

\institute{Instituto Galego de Física de Altas Enerxías IGFAE,
Universidade de Santiago de Compostela, E-15782 Galicia-Spain 
\and
           Faculty of Science and Technology, University of Stavanger, 4036 Stavanger, Norway 
\and
           Institute for Theoretical Physics, TU Wien, Wiedner Hauptstrasse 8-10, 1040 Vienna,
Austria
\and 
MIT Center for Theoretical Physics - a Leinweber Institute, Massachusetts Institute of Technology, Cambridge, MA 02139, USA
          }

\abstract{The effective kinetic theory (EKT) of QCD provides a possible picture of various non-equilibrium processes in heavy- and light-ion collisions. While there have been substantial advances in simulating the EKT in simple systems with enhanced symmetry, eventually, event-by-event simulations will be required for a comprehensive phenomenological modeling. As of now, these simulations are prohibitively expensive due to the numerical complexity of the Monte Carlo evaluation of the collision kernels. In this talk, we show how the evaluation of the collision kernels can be performed using neural networks paving the way to full event-by-event simulations.
}

\begin{flushright}
MIT-CTP/5932
\end{flushright}
\maketitle
\section{Introduction}
\label{intro}
Heavy-ion collisions at experiments at the LHC have been shown to produce one of the most extreme states of matter: the Quark-Gluon Plasma. This extremely short-lived fluid is believed to thermalize very quickly. A state-of-the-art description of this thermalization process is given by the so-called Effective Kinetic Theory (EKT), which captures the leading-order QCD scattering and splitting/merging processes~\cite{Arnold:2002zm}.

In its most general form, the microscopic behaviour of the system is governed by the leading order QCD Boltzmann equation,
\begin{equation}
    \left( \partial_t + \bf v \cdot \nabla_{\mathbf{x}} \right) f({\bf x}; {\bf p}; {\bf t}) = C_{1\leftrightarrow2}[f] + C_{2\leftrightarrow2}[f]~.
    \label{eq:BE}
\end{equation}
In this talk, we focus on the evolution of a pure gluon system; therefore, $f$ is the gluonic distribution function. Eq.~\eqref{eq:BE} describes the time evolution of an out-of-equilibrium system, where $1\leftrightarrow2$ and $2\leftrightarrow2$ processes are the underlying interactions of the partons that form the plasma.

Solving Eq.~\eqref{eq:BE} involves a numerical implementation in 3+3+1D, which is impractical from the computational point of view due to the large time required to compute the collision kernels. For this reason, different approximations have been used to study the thermalization of the Quark-Gluon Plasma. For instance, a longitudinal boost-invariant system with infinite transverse size~\cite{PhysRevD.27.140, Kurkela:2015qoa}, the diffusion approximation of the $2\leftrightarrow2$ kernel~\cite{Mueller:1999pi}, or Relaxation Time Approximation (RTA)~\cite{Kurkela_2020} are some examples that have been used to perform phenomenological studies of the hydrodynamization process.

We propose a novel approach based on Artificial Neural Networks (ANNs) to overcome the bottleneck of computing the collision kernels in less symmetric systems. Our approach exploits the fact that the QCD Boltzmann equation is local in space and, therefore, obtaining the collision kernels in each spatial cell requires doing a very similar calculation in each of them. Thus, this problem is suitable to be dealt with by training a neural network that fits the collision kernel for a given distribution function at every given spatial point. The locality will allow the application of the same ANN to each spatial cell. A more detailed explanation of the method we present here can be found in our recent paper~\cite{BarreraCabodevila:2025ogv}.

\section{Training data}

The first thing we need to take care of is which data we use to train our neural network. Since we want to create a map of a distribution function to its corresponding collision kernel, it is clear that both of them must be the input and output of the ANN, respectively. The calculation of the collision kernels is done by the well-known Monte Carlo solver of the EKT. We should restrict ourselves to physically sensible distributions, relevant for the thermalization process. Thus, we generate pairs of data with distribution functions corresponding to the kinetic evolution starting from initial conditions inspired by the CGC framework~\cite{Kurkela:2015qoa} and their respective collision kernels. Additionally, we will also include perturbations over the equilibrium distribution, which helps the network to approximate the collision kernels around thermal equilibrium.

To reduce the size of the data needed, we also exploit symmetries preserved by the Boltzmann equation. First, we take advantage of the conformal symmetry to fix the energy density of the distribution functions used in the training dataset. If we want to input a distribution with a different energy density, we can apply the corresponding conformal transformation such that the distribution has the required energy density. Then, after calculating the collision kernel, we perform the inverse transformation. Similarly, in the case of a 3D distribution function in momentum space, we establish a hierarchy for the anisotropies, such that the pressures are ordered $P_z>P_y>P_x$. Then, as before, if this is not true for the input distribution function, we perform a rotation before passing it to the ANN and rotate back the output.

This dataset needs to be preprocessed to improve the convergence of the subsequent training. Here, we enumerate the transformations we apply to the dataset:
\begin{itemize}
    \item Instead of using $f$ and $C$, we use the energy distributions, $p^3f$ and $p^3C$. This helps the network to have a good energy density conservation, which is a feature implemented in the Monte Carlo solver.
    \item We subtract the thermal equilibrium from the input since we observe that then the network reproduces better the thermal fixed point. Then, the mapping we fit is $p^3f - p^3f_{eq} \to p^3 C$.
    \item We standardize the training data, that is, we normalize the data such that each feature has a standard deviation equal to unity.
\end{itemize}
In total, we generate of the order of 100000 training pairs for both 1D system ($\sim100$ MB) and 3D system ($\sim 50$ GB).

\section{Choosing the ANN architecture}

The choice of the architecture of the neural network is, in principle, arbitrary. In our case, we restrict to neural networks with linear hidden layers and an arbitrary number of nodes per layer that use ReLU as activation function. To choose the optimal number of both internal layers and nodes, as well as the learning rate, we use RayTune~\cite{liaw2018tune}. This tool trains several networks in parallel and compares their performance to identify the better-performing ones.

We perform this procedure for $C_{1\leftrightarrow2}$ and $C_{2\leftrightarrow2}$ independently, such that we have two different networks to compute the collision kernels separately. Besides, we keep the 10 best-performing networks and not just the best one, so we can compare their output when presenting the results.

\section{Results}

The results we show here are obtained with the ten best fitted neural networks for each collision kernel. To compute the time evolution of a given initial distribution function, we produce ten independent evolutions with a fourth-order Runge-Kutta algorithm, one for each of the networks. Then, we show the mean value of the ten independent evolutions and estimate the error bands with the Jackknife method,
\begin{equation}
    \delta f(t_n) = \sqrt{\frac{M-1}{M} \sum_m \left( f_{(m)}(t_n) - \langle f(t_n) \rangle \right)^2}~.
    \label{eq:JK}
\end{equation}
The trained neural networks, as well as the training dataset used for the 1D case, are publicly available in~\cite{zenododataset}. In the following, let us briefly comment on the results we have obtained.

\begin{figure}
    \centering
    \includegraphics[width=0.8\linewidth]{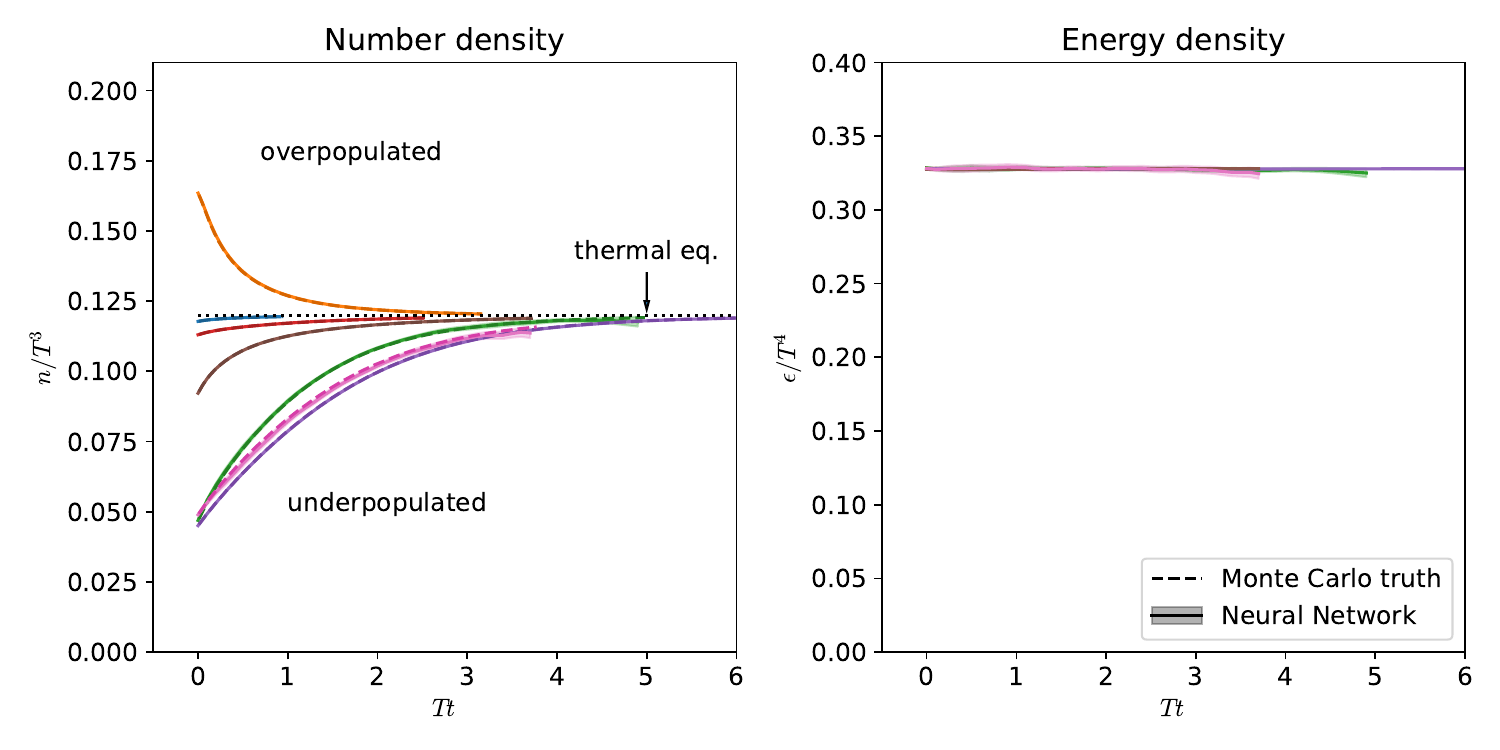}
    \caption{Number (left) and energy (right) densities time evolutions for different initial conditions in the isotropic/1D case. Solid lines with error bars correspond to the neural network predictions, while the dashed lines are computed with the Monte Carlo algorithm. Figure obtained from~\cite{BarreraCabodevila:2025ogv}.}
    \label{fig:1D}
\end{figure}

First, let us focus on the results for the 1D scenario. In Fig.~\ref{fig:1D} we display the number and energy density for different evolutions. In all cases, the energy density is nicely conserved, and the number density approaches its thermal equilibrium value following the same trend as the Monte Carlo calculation. It is relevant to mention that when the system is close to equilibrium, the error bars start to grow and the evolution is not fully stable.

\begin{figure}
    \centering
    \includegraphics[width=0.38\linewidth]{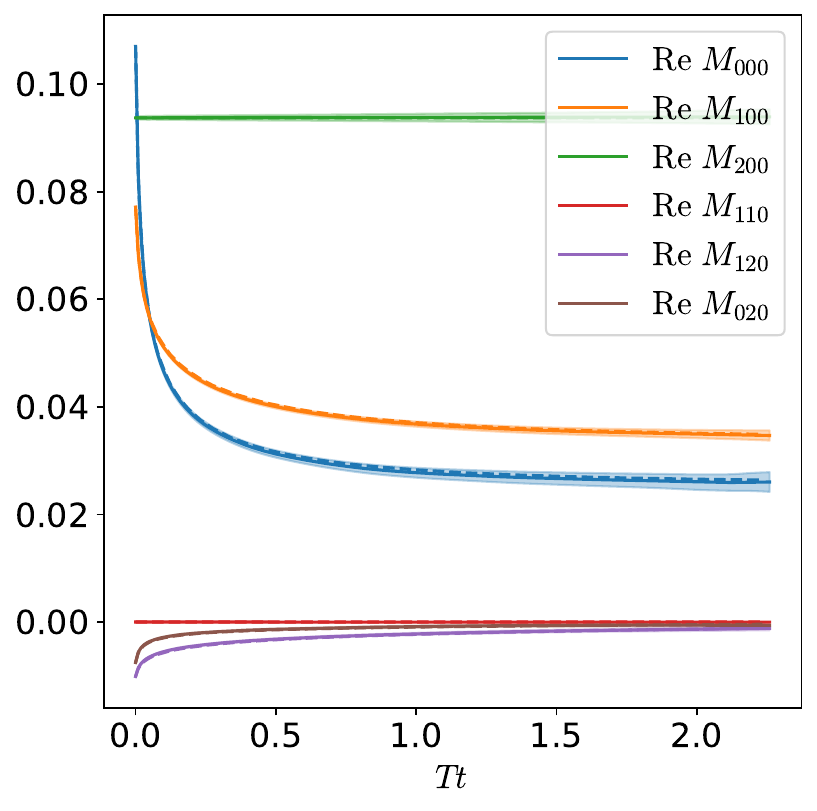}
    \includegraphics[width=0.38\linewidth]{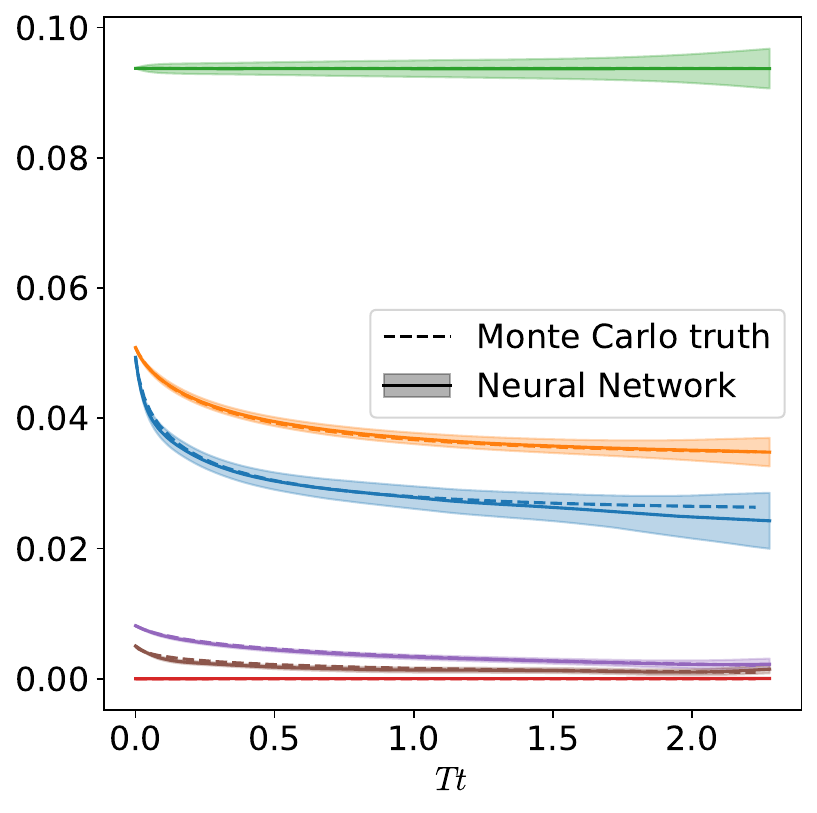}
    \caption{Different moments of the distribution defined in Eq.~\eqref{eq:momentsadim} as a function of time for two different initial conditions. Figure obtained from~\cite{BarreraCabodevila:2025ogv}.}
    \label{fig:3D}
\end{figure}

Regarding the 3D scenario, we show moments of the distribution, defined as
\begin{align}
    M_{nlm}=\frac{1}{T^{n+2}}\int\frac{d^3{\bf p}}{(2\pi)^3}p^{n-1} Y^m_l{}^\ast(\theta,\phi)f({\bf p}),
    \label{eq:momentsadim}
\end{align}
in Fig.~\ref{fig:3D}. In this expression, $Y^m_l(\theta,\phi)$ are the spherical harmonics. As in the previous case, the curves follow the same trend as the results obtained in the Monte Carlo approach. The energy density, given by the $M_{200}$ moment, is conserved, and at later times, the error bars grow, indicating that the network has challenges in capturing the thermal fixed point.

A detailed benchmark comparison between the Monte Carlo and the novel neural network approaches is subtle and depends a lot on the chosen grid and the desired accuracy. However, as an estimate, we observe a systematic speed-up of roughly three orders of magnitude in the calculation of the full evolution, making it an extremely attractive method despite its challenges near equilibrium.

\section*{Acknowledgements}

FL is a recipient of a DOC Fellowship of the Austrian Academy of Sciences at TU Wien (project 27203). This work is funded in part by the Austrian Science Fund (FWF) under Grant DOI 10.55776/P34455, and Grant DOI 10.55776/J4902. For the purpose of open access, the authors have applied a CC BY public copyright license to any Author Accepted Manuscript (AAM) version arising from this submission. The results in this paper have been achieved in part using the Austrian Scientific Computing (ASC) infrastructure, project 71444. SBC is supported by the European Research Council project ERC-2018-ADG-835105 YoctoLHC; by Mar\'\i a de Maeztu grant CEX2023-001318-M and by project PID2023-152762NB-I00, both funded by MCIN/AEI/10.13039/-501100011033; from the Xunta de Galicia (CIGUS Network of Research Centres) and the European Union.

%
\bibliography{bibliography.bib} 
%
%
%
%


\end{document}